\documentclass[conference, a4paper]{IEEEtran}
\usepackage[utf8]{inputenc}
\usepackage{lipsum}
\usepackage{amsmath,amssymb,amsthm}
\usepackage[linesnumbered,ruled,vlined,titlenumbered]{algorithm2e}
\usepackage[innermargin=0.545in,outermargin=0.545in,top=0.545in,bottom=.7in]{geometry}
\usepackage{algorithmicx}
\usepackage[colorlinks=true,citecolor=blue,linkcolor=blue,urlcolor=blue]{hyperref}

\usepackage{cite}

\usepackage{tikz}
\usepackage{pgfplots}

\usepackage{tikzscale}
\pgfplotsset{compat=newest}
\usepackage{xcolor}
\usepackage{mathtools}
\usepackage{url}
\usepackage{booktabs}

\newtheorem{theorem}{Theorem}
\newtheorem{definition}{Definition}
\newtheorem{notation}{Notation}
\newtheorem{lemma}{Lemma}

\newtheorem{remark}{Remark}

\def\ve#1{{\mathchoice{\mbox{\boldmath$\displaystyle #1$}}%
		{\mbox{\boldmath$\textstyle #1$}}%
		{\mbox{\boldmath$\scriptstyle #1$}}%
		{\mbox{\boldmath$\scriptscriptstyle #1$}}}}

\IEEEoverridecommandlockouts

\author{\IEEEauthorblockN{Cornelia Ott$^{1}$, Sven Puchinger$^{2}$, Martin Bossert$^{1}$}
\IEEEauthorblockA{%
$^{1}$Institute of Communications Engineering, Ulm University, Germany \\
$^{2}$Department of Applied Mathematics and Computer Science, Technical University of Denmark (DTU), Denmark \\
{\small E-mail: cornelia.ott@uni-ulm.de, svepu@dtu.dk, martin.bossert@uni-ulm.de}
}

}
\title{
Bounds and Genericity of Sum-Rank-Metric Codes
}

\newcommand{\Fq}{\mathbb{F}_q}
\newcommand{\Fqm}{\mathbb{F}_{q^m}}
\newcommand{\NN}{\mathbb{N}}

\newcommand{\Code}{\mathcal{C}}
\newcommand{\x}{\ve{x}}
\renewcommand{\c}{\ve{c}}
\newcommand{\G}{\ve{G}}
\newcommand{\I}{\ve{I}}
\newcommand{\X}{\ve{X}}
\renewcommand{\H}{\ve{H}}
\newcommand{\0}{\ve{0}}
\newcommand{\A}{\ve{A}}
\newcommand{\U}{\ve{U}}
\newcommand{\y}{\ve{y}}
\newcommand{\info}{\ve{i}}
\newcommand{\wtr}{\mathrm{wt_{Rk}}}
\newcommand{\wtsr}{\mathrm{wt}_{SR,\ell}}
\newcommand{\dsr}{\mathrm{d}_{SR,\ell}}
\newcommand{\Sl}{\mathcal{S}_\ell}
\newcommand{\Bl}{\mathcal{B}_\ell}
\newcommand{\VolS}{\mathrm{Vol}_{\mathcal{S}_{\ell}}}
\newcommand{\VolB}{\mathrm{Vol}_{\mathcal{B}_{\ell}}}
\newcommand{\rk}{\mathrm{rk}}

\makeatletter
\newcommand{\removelatexerror}{\let\@latex@error\@gobble}
\makeatother

\usetikzlibrary{arrows,matrix,positioning,patterns}
\definecolor{constructionEcolor}{rgb}{0,0.7,0}	

\usepackage{ifthen}
\newboolean{ShortVersion}
\setboolean{ShortVersion}{false}

\IEEEoverridecommandlockouts
\begin{document}

\maketitle
\thispagestyle{empty}
\pagestyle{empty}

\begin{abstract}
We derive simplified sphere-packing and Gilbert--Varshamov bounds for codes in the sum-rank metric, which can be computed more efficently than previous ones.
They give rise to asymptotic bounds that cover the asymptotic setting that has not yet been considered in the literature:
families of sum-rank-metric codes whose block size grows in the code length.
We also provide two genericity results: we show that random linear codes achieve almost the sum-rank-metric Gilbert--Varshamov bound with high probability.
Furthermore, we derive bounds on the probability that a random linear code attains the sum-rank-metric Singleton bound, showing that for large enough extension field, almost all linear codes achieve it.
\end{abstract}

\section{Introduction}

The sum-rank metric is a mix of the Hamming and rank metric.
It was first introduced in 2010 \cite{nobrega2010multishot}, motivated by multi-shot network coding.
Since then, many code constructions and decoding algorithms for sum-rank-metric codes have been proposed \cite{wachter2011partial,wachter2012rank,wachter2015convolutional,napp2017mrd,napp2018faster,martinez2018skew,boucher2019algorithm,martinez2019reliable,caruso2019residues,bartz2020fast,martinezpenas2020sumrank,byrne2020fundamental}. 
Some of these codes have found applications in distributed storage \cite{martinez2019universal}, further aspects of network coding \cite{martinez2019reliable}, and space-time codes \cite{shehadeh2020rate}.

In two extreme cases, the metric coincides with the Hamming and the rank metric, respectively, and thus many fundamental bounds on the code parameters are known 
\cite{hamming1950error,gilbert1952comparison,varshamov1957estimate,loidreau2006properties} for these two cases.
Although the sum-rank metric has been studied since 2010, only very recently, Byrne, Gluesing-Luerssen, and Ravagnani \cite{byrne2020fundamental} presented (among many other fundamental results) a sphere-packing and Gilbert--Varshamov bound for sum-rank metric codes.
They also presented asymptotic versions of the bounds on sum-rank metric codes for bounded block sizes and growing number of blocks.
The bounds for finite parameters depend on the sum-rank-metric ball size, which is super-polynomial to compute using the presented formula.

Furthermore, it is well-known that random codes in the Hamming and rank metric \cite{loidreau2006properties}
achieve the respective Gilbert--Varshamov bound with high probability, hence codes attaining these bounds are the generic case.
Bounds on the probability that random codes fulfill the Hamming or rank-metric Singleton bound with equality (called maximum distance separable (MDS) or maximum rank distance (MRD) codes, respectively), have been derived in \cite{ravagnani2019sparsity}
and \cite{neri2018genericity}, respectively. No such result is known for the sum-rank metric, where codes attaining the Singleton bound \cite{martinez2018skew} are called maximum sum-rank distance (MSRD) codes.

In this paper, we extend these the results by Byrne et al., as well as the genericity results from the Hamming and rank metric, as follows.
We present variants of the sphere-packing (SP) and Gilbert--Varshamov (GV) bound for linear codes and draw the connection to a recent algorithm to compute sum-rank-metric sphere sizes \cite{puchinger2020generic}, which allows to compute the bounds in polynomial time.
Using lower and upper bounds on the sum-rank-metric ball size, we derive simplified bounds that can be computed even more efficiently.
These simplified bounds also induce asymptotic variants of the two bounds, 
which extend the results in Byrne et al.\ by covering also the case of growing block size.

Furthermore, we present the following genericity results: we show that random linear codes achieve almost the sum-rank GV bound with high probability and derive two bounds on the probability that a random code is MSRD. The bounds smoothly interpolate between the known bounds in the Hamming and rank metric and show that MSRD codes are generic for growing extension degree of the underlying field.

\ifShortVersion
Due to space limitation, some proofs and details are omitted, but can be found in an extended version of the paper~\cite{ott2021extended}.
\fi

\section{Preliminaries}

We use a similar notation as in \cite{puchinger2020generic}.
Let $q$ be a prime power and $m, n, \ell, \eta$ positive integers. We denote by $\Fq$ a finite field with $q$ elements and by $\Fqm$ its extension field. The codes we consider in this paper are subsets of $\Fqm^n$,
where each vector $\x=[\x_1| \ldots| \x_\ell] \in \Fqm^n$ consists of $\ell$ blocks $\x_1, \ldots, \x_\ell \in \Fqm^\eta$ of length $\eta$. Therefore we assume $n=\ell\cdot \eta$. 
Since $\Fqm$ is a a vectorspace over $\Fq$ of dimension $m$, a vector $\x_i \in \Fqm^\eta$ can also be represented as a matrix $\ve{X}_i \in \Fq^{m \times \eta}$, hence the rank weight of $\x_i$ is defined as $\wtr(\x_i) \coloneqq \dim_{\Fq} \langle x_1, \ldots, x_\eta\rangle$, which is equal to the rank of the matrix $\ve{X}_i$. Clearly it holds  for $\x_i \in \Fqm^\eta$ that $\wtr(\x_i)\in \{0, \ldots, \mu\}$, where $\mu\coloneqq \min\{m,\eta\}$. We define sum-rank weight and the sum-rank distance of a vector $\x\in \Fqm^n$ as follows.

\begin{definition}
Let $\x=[\x_1| \ldots| \x_\ell] \in \Fqm^n$. \emph{The ($\ell$-)sum rank weight} of $\x$ is defined as 
\[
\wtsr: \Fqm^n \rightarrow \NN, \quad \x \mapsto \textstyle\sum_{i=0}^{\ell}\wtr(\x_i).
\]
For two vectors $\x, \x' \in \Fqm^n$ the \emph{($\ell$-)sum rank distance} is defined as 
\begin{align*}
\dsr: \Fqm^n \times \Fqm^n &\rightarrow  &&\NN, \\ (\x,\x') &\mapsto  &&\dsr(\x,\x')\coloneqq\wtsr(\x-\x').
\end{align*}
The vector $[\wtr(\x_1), \ldots, \wtr(\x_\ell)]$ is called the \emph{weight decomposition} of $\x$.
\end{definition}
The ($\ell$-)sum-rank distance $\dsr$ is a metric over $\Fqm^n$, the socalled \emph{sum-rank metric}. In the following we define spheres and balls in the sum-rank metric analogues to \cite{loidreau2006properties} and give defintions for their volume.

\begin{definition}
Let $\tau \in \mathbb{Z}_{\geq 0}$ with $0\leq\tau\leq \ell\cdot \mu$ and $\x \in \Fqm^n$.
The sum-rank-metric sphere with radius $\tau$ and center $\x$ is defined as
\[
\Sl(\x,\tau)\coloneqq \{\y\in \Fqm^n \mid \dsr(\x,\y)=\tau\}.
\]
Analogously, we define the ball of sum-rank radius $\tau$ with center $\x$ by
\[
\Bl(\x,\tau) \coloneqq \textstyle\bigcup_{i=0}^{\tau}\Sl(\x,i).
\]
We also define the following cardinalities:
\begin{align*}
\VolS(\tau) &\coloneqq |\{\y\in \Fqm^n \mid \wtsr(\y)=\tau\}|, \\
\VolB(\tau) &\coloneqq \textstyle\sum_{i=0}^{\tau}\VolS(i).
\end{align*}
\end{definition}

Since the sum-rank metric is invariant under translation of vectors, the volume of a sphere or ball is independent of its center. Hence, $\VolS(\tau)$ and $\VolB(\tau)$ are the volumes of any sphere or ball of radius $\tau$.
Unlike the extreme cases, Hamming and rank metric, it is quite involved to compute these volumes in general. The formula given in \cite{byrne2020fundamental} consists of a sum whose number of summands may grow super-polynomially in $\tau$, depending on the relative size of $\ell$ and $\eta$.
In \cite{puchinger2020generic}, a dynamic-programming algorithm was given, which computes the volumes in polynomial time.

We define a linear sum-rank metric code as follows.
\begin{definition}
A linear sum-rank metric code $\Code$ over $\Fqm$ of length $n$ and dimension $k$ is an $\Fqm$-vector space $\Code\subset \Fqm^n$ with $\dim_{\Fqm}(\Code)=k$. Hence, the cardinality of the code is $|\Code|=q^{mk}$. Each codeword $\c=[\c_0|\ldots| \c_\ell]\in \Code$ consists of $\ell$ blocks $\c_i\in \Fqm^\eta$ of length $\eta$. The minimum ($\ell$)-sum-rank distance $d$ is defined as
\[
d \coloneqq \min_{\c\neq \c'\in \Code}\{\dsr(\c,\c')\}=\min_{\c \in \Code}\{\wtsr(\c)\}.
\]
We denote such a code by $\Code(n,k,d)$.
\end{definition}
The sum-rank weight of a vector $\x\in \Fqm^n$ is at most its Hamming weight.
This implies the following Singleton bound in the sum-rank metric.
\begin{theorem}[\!\!{\cite[Proposition 34]{martinez2018skew}}]
Let $\Code(n,k,d)$ be a linear sum-rank metric code. Then it holds
$
d\leq n-k+1.
$
\end{theorem}
Codes that fulfill this bound with equality are called \emph{maximum sum-rank distance codes} (MSRD codes). In \cite[Theorem 4]{{martinez2018skew}} it is shown, that the therein defined \emph{Linearized Reed--Solomon (LRS) codes} are MSRD codes. The code parameters of LRS codes are restricted by $\ell <q$ and $\eta \leq m$. It is particularly interesting to know bounds on the code parameters for cases in which these restrictions are not met.

We define the set
\[
\tau_{t,\ell,\mu}\coloneqq\left\{\ve{t}=(t_1, \ldots, t_\ell)\mid \textstyle\sum_{i=1}^\ell t_i =t, t_i\leq \mu\quad \forall i\right\},
\]
which corresponds combinatorially to the set of ordered partitions with bounded number of summands and bounded summands.
We will extensively use the number of such partitions throughout the paper.
By common combinatorical methods, we get
\begin{align}
\label{ineq: tau}
|\tau_{t,\ell,\mu}| = \textstyle\sum_{i=0}^{\lfloor\frac{t}{\mu+1}\rfloor} (-1)^i \tbinom{\ell}{i}\tbinom{t+\ell-1-(\mu+1)i}{\ell-1} \leq \tbinom{t+\ell-1}{\ell-1}
\end{align}
(see also \cite[Lemma 1.1]{ratsaby2008estimate}). The upper bound $\binom{t+\ell-1}{\ell-1}$ can also be easily derived by a stars-$\&$-bars argument.

\section{Bounds in sum-rank metric}

In this section, we present bounds on sum-rank-metric codes.
The first two subsections contain slight reformulations, for the case of linear codes, of the SP and GV bounds presented in \cite{byrne2020fundamental}.
We also state the (polynomial) complexity of computing the bounds if the efficient dynamic-programming method in \cite{puchinger2020generic} is used to compute $\VolS(\tau)$, instead of the formula in \cite{byrne2020fundamental}, which has super-polynomially many summands.
The main results of this section are the simplified and asymptotic SP and GV bounds in Section~\ref{ssec:simplified_and_asymptotic_bounds}, which we derive from upper and lower bounds on $\VolS(\tau)$.
We conclude the section with numerical comparisons of the bounds.

\subsection{Sphere-Packing Bound}

We give an SP bound for linear codes in sum-rank metric by specializing the argument in \cite{byrne2020fundamental} to linear codes.
\begin{theorem}\label{SP_bound}
For a linear sum-rank metric code $\Code(n,k,d)$, it holds that
\[
q^{mk}\cdot \VolB\Big(\Big\lfloor\frac{d-1}{2}\Big\rfloor\Big)\leq q^{mn}.
\]
Furthermore, both sides of the bound can be computed in complexity ${\mathcal{O}}^{\sim}\big(\ell^2d^3+\ell d^4(m+\eta)\log(q)\big)$  using the efficient algorithm for computing $|\VolS|$ in \cite[Theorem 5 and Algorithm 1]{puchinger2020generic}.
\end{theorem}
\ifShortVersion
\begin{IEEEproof}
See extended version of this paper~\cite{ott2021extended}.
\end{IEEEproof}
\else
\begin{IEEEproof}
Since the minimum sum-rank distance of $\Code$ is $d$ it holds for all $\c_1, \c_2 \in \Code$ with $\c_1\neq\c_2$ that $\Bl(\c_1,\lfloor\frac{d-1}{2}\rfloor)\cap\Bl(\c_2,\lfloor\frac{d-1}{2}\rfloor)=\emptyset$ and
hence
\[\Big|\bigcup\textstyle_{\c\in\Code}\Bl\Big(\c,\Big\lfloor\frac{d-1}{2}\Big\rfloor\Big)\Big|=\textstyle\sum_{\c\in\Code}\Big|\Bl\Big(\c,\Big\lfloor\frac{d-1}{2}\Big\rfloor\Big)\Big|.
\]
With this fact the relation
\[
\bigcup\textstyle_{\c\in\Code} \Bl\
\Big(\c, \Big\lfloor\frac{d-1}{2}\Big\rfloor\Big)\subseteq \Fqm^n
\]
leads to
\[
\Big|\bigcup\textstyle_{\c\in\Code} \Bl\
\Big(\c, \Big\lfloor\frac{d-1}{2}\Big\rfloor\Big)\Big|=|\Code|\cdot\VolB\Big(\Big\lfloor\frac{d-1}{2}\Big\rfloor\Big)\leq |\Fqm^n|.
\]
The heaviest computational step for evaluating the bound is to determine $\VolB$, which can be done by computing the sphere size $\VolS(\tau)$ for $\tau=0,\dots,\lfloor\frac{d-1}{2}\rfloor$. This can be done in the claimed complexity by calling Algorithm~1 in \cite{puchinger2020generic} at most $\lfloor\frac{d-1}{2}\rfloor+1$ times.
\end{IEEEproof}
\fi
\subsection{A Gilbert--Varshamov like bound}
In this subsection we derive a pendant to the GV bound for the sum-rank metric for linear codes $\Code\subset \Fqm^n$ of length $n$ and dimension $k$.
The statement is slightly different than the GV bound in \cite{byrne2020fundamental}: we show the existence of a \emph{linear} code instead of an arbitrary code.

\begin{theorem}[Gilbert--Varshamov bound]\label{GV-Bound}
Let $\Fqm$ be a finite field, $\ell, n, k, d\leq \mu \ell$ be positive integers that satisfy 
\begin{align}
\label{ineq: GV}
q^{m(k-1)}\cdot \VolB(d-1)< q^{mn}.
\end{align}
Then, there is a linear code of length $n$, dimension $k$, and minimum $\ell$-sum-rank distance at least $d$.
As in Theorem~\ref{SP_bound}, we can compute both sides of the bound in complexity ${\mathcal{O}}^{\sim}\big(\ell^2d^3+\ell d^4(m+\eta)\log(q)\big)$  using the efficient algorithm for computing $|\VolS|$ in \cite[Theorem 5 and Algorithm 1]{puchinger2020generic}.
\end{theorem}
\begin{IEEEproof}
We consruct a linear code of length $n$, minimum sum-rank distance $d$ and dimension $k=1$. Let $\c_0\coloneqq [0\ldots0]\in \Fqm^n$ and let $\c_1\in\Fqm^n$ with $\wtsr(\c_1)=d$. Let $\Code\coloneqq \langle \c_0, \c_1 \rangle$ then it holds that $|\Code|=|\{\c_0, \alpha \c_1\mid \forall \alpha \in \Fqm\setminus\{0\}\}|=q^m$. One can see, that $\wtsr(\c_0+\alpha\c_1)=d$ for all $\alpha \in \Fqm$.

Inductively, we assume that we have a linear code $\Code(n,k-1,d)\subset \Fqm^n$ fulfilling (\ref{ineq: GV}).
Since $\bigcup_{\c\in\Code}\Bl(\c,d)\subsetneq \Fqm^n $ we can choose a vector $\c'\in  \Fqm^n \setminus \bigcup_{\c\in\Code}\Bl(\c,d)$. 
Obviously it holds that $\dsr(\c, \c')\geq d \quad\forall \c \in \Code$. 
We contruct now a vector space $\Code'\coloneqq \langle \Code \cup \c'\rangle=\{\c+ \alpha \c'\mid \forall \alpha \in \Fqm\, \c \in \Code\}$. For all $\c_0 \in \Code$, $\c_1+\alpha \c' \in \Code'$ with $\alpha \neq 0$ it holds 
$
\dsr(\c_0,\c_1+\alpha \c')=\wtsr(\alpha^{-1}(\c_1-\c_0)+ \c').
$
Since $\c\coloneqq\alpha^{-1}(\c_1-\c_0)\in \Code$ one get
$
\wtsr(\c+\c')=\dsr(\c, \c')\geq d.
$
The complexity result follows from \cite[Theorem 5]{puchinger2020generic}.
\end{IEEEproof}

\subsection{Simplified and Asymptotic Bounds}\label{ssec:simplified_and_asymptotic_bounds}

In this section, we derive simplified versions of the SP and GV bound based on lower and upper bounds on the volume of a sum-rank-metric ball.
These simplified bounds immediately give new asymptotic bounds for the two cases, in asymptotic settings for which no asymptotic bounds are known.

In order to give a lower bound on $\VolB(t)$, we first derive in the following lemma a lower bound for the number of $m\times n$ matrices over $\Fq$ for a given rank $t \leq \min\{m,n\}$ which is denoted by $\mathrm{NM}_q(n,m,t)$.
The exact number of $\mathrm{NM}_q(n,m,t)$ was given in \cite{migler2004weight}:
\[
\mathrm{NM}_q(n,m,t) = \begin{bmatrix}n\\t\end{bmatrix}_{q} \cdot \prod\textstyle_{i=0}^{t-1}(q^m-q^i).
\]
We define
\begin{equation}
\gamma_q := \prod\textstyle_{i=1}^{\infty} (1-q^{-i})^{-1}. \label{eq:gamma_q}
\end{equation}
Note that $\gamma_q$ is monotonically decreasing in $q$ with a limit of $1$, and e.g.~$\gamma_2 \approx 3.463$, $\gamma_3 \approx 1.785$, and $\gamma_4 \approx 1.452$.
\begin{lemma} The cardinality of all  $m\times n$ matrices over $\Fq$ of rank $t \leq \min\{m,n\}$ is bounded by
\[\mathrm{NM}_q(n,m,t)\geq q^{(m+n-t)t}\gamma_q^{-1},\]
with $\gamma_q \leq 3.5$ defined as in \eqref{eq:gamma_q}.
\end{lemma}
\begin{IEEEproof}
The $q$-binomial coefficient is denoted by $\begin{bmatrix}n\\k
\end{bmatrix}_{q}$ and here $q$ is a prime power.
In \cite{koetter2008coding} the following lower bound for the $q$-binomial coefficient was given:
$
\begin{bmatrix}n\\t\end{bmatrix}_{q}\geq q^{(n-t)t}.
$
Therefore we get 
\begin{align*}
\mathrm{NM}_q(n,m,t) &\geq q^{(n-t)t}\textstyle\prod_{i=0}^{t-1}(q^m-q^i)\\
&\geq q^{(n+m-t)t}\textstyle\prod_{j=1}^{t}(1-q^{-j})
\geq q^{(n+m-t)t}\gamma_q^{-1}. 
\end{align*}
\end{IEEEproof}
Using this bound allows us to give a lower bound on the volume of a sphere $\VolS(t)=\sum_{\ve{t}\in \tau_{t,\ell,\mu}}\prod_{i=1}^\ell \mathrm{NM}_q(\eta,m,t_i)$ containing all vectors in $\Fqm^n$ of sum-rank weight $t$.

\begin{lemma}\label{Vol_lower_bounds}
For the volumes of a sphere and of a ball with sum-rank radius $t$ it holds:
\[
\VolB(t)\geq\VolS(t)\geq q^{(m+\eta-\frac{t}{\ell})t-\frac{\ell}{4}}\cdot \gamma_q^{-\ell}.
\]
\end{lemma}

\begin{IEEEproof}
We have
\begin{align*}
\VolS(t)&=\textstyle\sum_{\ve{t}\in \tau_{t,\ell,\mu}}\prod_{i=1}^\ell \mathrm{NM}_q(\eta,m,t_i)\\
&\geq\max_{\ve{t}\in \tau_{t,\ell,\mu}}\Big\{\prod\textstyle_{i=1}^\ell q^{(m+\eta-t_i)t_i}\gamma_q^{-1}\Big\}.\\
&=q^{(m+\eta)t}\cdot q^{-\min_{\ve{t}\in \tau_{t,\ell,\mu}}\{\sum_{i=1}^\ell t_i^2\}}\cdot \gamma_q^{-\ell}.
\end{align*}
We can write $t$ as $t=t_*\cdot \ell +r$, with $0\leq r <\ell$.
Since the expression $\sum_{i=1}^\ell t_i^2$ is minimized by the quasi-equal decomposition:
$\ve{t}=(t_1,\ldots, t_\ell)$ with $t_{i_1}=\ldots=t_{i_r}=t_*+1$ and $t_{i_r}=\ldots=t_{i_l}=t_*$, we get
\begin{align*}
\min_{\ve{t}\in \tau_{t,\ell,\mu}}\Big\{\sum_{i=1}^\ell t_i^2\Big\}
&= r \cdot (t_{*}+1)^2+(\ell-r)\cdot t_*^2
=\frac{t^2-r^2}{\ell}+r.
\end{align*}
Since  $\max_{r\in \mathbb{N}_{\leq\ell-1}}\{r-\frac{r^2}{\ell}\}\leq\frac{\ell}{4}$ it holds that $\min_{\ve{t}\in \tau_{t,\ell,\mu}}\Big\{\textstyle\sum_{i=1}^\ell t_i^2\Big\} \leq \frac{t^2}{\ell}+\frac{\ell}{4}$ and hence
$\VolS(t)\geq q^{(m+\eta)t}\cdot q^{-\frac{t^2}{\ell}-\frac{\ell}{4}}\cdot \gamma_q^{-\ell}.$
Since the volume of a ball is always greater than the volume of a sphere with the same radius, the statement follows.
\end{IEEEproof}

\begin{remark}
It can be seen from the proof of Lemma~\ref{Vol_lower_bounds} that for $\ell \mid t$, we have
\begin{align*}
\VolS(t)\geq q^{(m+\eta-\frac{t}{\ell})t}\cdot \gamma_q^{-\ell},
\end{align*}
i.e., we can drop the term $-\frac{\ell}{4}$ in the exponent of $q$.
\end{remark}

\begin{theorem}[Simplified SP Bound]\label{Simplified_SP_Bound}
For a linear sum-rank metric code $\Code(n,k,d)$, the parameters fulfill
\[
q^{mk}\cdot q^{(m+\eta-\frac{1}{\ell}\lfloor \frac{d-1}{2}\rfloor)\lfloor \frac{d-1}{2}\rfloor-\frac{\ell}{4}}\cdot \gamma_q^{-\ell} \geq q^{mn}.
\]
\end{theorem}
\ifShortVersion
\begin{IEEEproof}
See extended version of this paper~\cite{ott2021extended}.
\end{IEEEproof}
\else
\begin{IEEEproof}
This follows directly fom Theorem~\ref{SP_bound} and Lemma~\ref{Vol_lower_bounds}.
\end{IEEEproof}
\fi

\begin{theorem}[Asymptotic SP Bound]
\label{Theo:asy_SP}
Let $\Code(n,k,d)$ be a linear sum-rank metric code and $\delta\coloneqq \frac{d}{n}$ the relative minimum distance.
Then the code rate $\mathcal{R}=\frac{k}{n}$ is upper bounded by 
\begin{align*} 
\mathcal{R}&< \delta^2 \frac{\eta}{4m}-\delta\Big(\frac{1}{2}+\frac{\eta}{m}\Big(\frac{1}{2}+\frac{1}{n}\Big)\Big)+\frac{1}{n}\Big(1+\frac{\eta}{m}+\frac{\eta}{nm}\Big)\\
&\quad +\frac{1}{\eta m}\Big(\frac{1}{4}+\log_q(\gamma_q)\Big)+1=:\mathcal{R}^*(\delta).
\end{align*}
Let $\xi>0$ be fixed. Then,
\begin{itemize}
\item[(i)]
For $m=\eta\xi \rightarrow \infty$ we get $\mathcal{R}\sim
\delta^2 \frac{1}{4\xi}
-\frac{\delta}{2}\Big(1+\frac{1}{\xi}\Big)+1.
$
\item[(ii)]
For $\ell \rightarrow \infty$ one get $\mathcal{R}\sim
\delta^2 \frac{\eta}{4m}-\frac{\delta}{2}\Big(1+\frac{\eta}{m}\Big)\\
\quad+\frac{1}{\eta m}\Big(\frac{1}{4}+\log_q(\gamma_q)\Big)+1.
$
\end{itemize}
\end{theorem}
\ifShortVersion
\begin{IEEEproof}
See extended version of this paper~\cite{ott2021extended}.
\end{IEEEproof}
\else
\begin{IEEEproof}

We transform the simplified SP bound (cf.\ Theorem~\ref{Simplified_SP_Bound}) with $t\coloneqq \lfloor \frac{d-1}{2}\rfloor$ into
\begin{align*}
\frac{k}{n} \leq 1-\frac{\Big(m+\eta-\frac{t}{\ell}\Big)t-\ell \Big(\frac{1}{4}+\log_q(\gamma_q)\Big)}{mn}.
\end{align*}
With
$\delta=\frac{d}{n}$ using $t\geq\frac{d-2}{2}=\frac{1}{2}\delta n-1$, it follows
\begin{align*}
\frac{k}{n} \leq
&\delta^2 \frac{\eta}{4m}-\delta\Big(\frac{1}{2}+\frac{\eta}{m}\Big(\frac{1}{2}+\frac{1}{n}\Big)\Big)+\frac{1}{n}\Big(1+\frac{\eta}{m}+\frac{\eta}{nm}\Big)\\
&+\frac{1}{\eta m}\Big(\frac{1}{4}+\log_q(\gamma_q)\Big)+1=:\mathcal{R}^*(\delta).
\end{align*}
Let $\xi$ be a constant. Consider the following limits:
\begin{itemize}
\item[(i)]$
\lim_{m=\eta\xi \rightarrow \infty}\mathcal{R}^*(\delta)=
\delta^2 \frac{1}{4\xi}
-\frac{\delta}{2}\Big(1+\frac{1}{\xi}\Big)+1
$
\item[(ii)]$
\lim_{\ell \rightarrow \infty}\mathcal{R}^*(\delta)=
\delta^2 \frac{\eta}{4m}-\frac{\delta}{2}\Big(1+\frac{\eta}{m}\Big)\\
+\frac{1}{\eta m}\Big(\frac{1}{4}+\log_q(\gamma_q)\Big)+1.
$
\end{itemize}
\end{IEEEproof}
\fi
In a similar fashion, we derive a simplified GV bound, for which we rely on an upper bound on $\VolS(t)$, which was derived in \cite{puchinger2020generic}.
We assume $d>2$ to avoid a more technical statement.
\begin{theorem}[Simplified GV Bound]
\label{theo:simpl_GV}
Let $\Fqm$ be a finite field, $\ell, n, k,d $ be positive integers with $2<d\leq \mu \ell$ that satisfy 
\begin{align*}
q^{m(k-1)}\cdot (d-1)\binom{\ell+d-2}{\ell-1}\gamma_q^\ell q^{(d-1)(m+\eta-\frac{d-1}{\ell})}< q^{mn}.
\end{align*}
Then, there is a linear code of length $n$, dimension $k$, and minimum $\ell$-sum-rank distance at least $d$.
\end{theorem}
\ifShortVersion
\begin{IEEEproof}
See extended version of this paper~\cite{ott2021extended}.
\end{IEEEproof}
\else
\begin{IEEEproof}
In \cite[Theorem 4]{puchinger2020generic} the following upper bound on the sphere size $\VolS(t)$ was given: 
\begin{align*}
\VolS(t)\leq\binom{\ell+t-1}{\ell-1}\gamma_q^\ell q^{t(m+\eta-\frac{t}{\ell})},
\end{align*}
Due to $\VolB(t) = \sum_{t'=0}^{t} \VolS(t') \leq t\VolS(t)$ fot $t>1$, 
this gives an upper bound on $\VolB(t)$. Together with Theorem~\ref{GV-Bound}, the claim follows.
\end{IEEEproof}
\fi
\begin{theorem}[Asymptotic Gilbert--Varshamov-like Bound]
\label{Theo: asy GV} 
For a finite field $\Fqm$ and  positive integers $\ell, n, \mathcal{R}n,d $ with $\delta\coloneqq\frac{d}{n}$  and $2<d\leq \mu \ell$ satisfying
\begin{align*}
\mathcal{R}\leq&\delta^2 \frac{\eta}{m}-\delta\Big(1+\frac{\eta}{m}+\frac{2\eta}{nm}\Big)+1+\frac{1}{n}+\frac{\eta}{nm}+\frac{\eta}{n^2m}\\
&-\frac{\sum_{i=1}^{\delta n-1}\log_q\Big(1+\frac{\ell-1}{i}\Big)+\log_q(\delta n-1)}{mn} -\frac{\log_q(\gamma_q)}{\eta m}
\end{align*}
there exist  a linear $\ell$-sum-rank metric code of rate $\mathcal{R}$ and relative minimum sum-rank distance at least $\delta$.
Let $\xi$ be a constant. For $m=\eta\xi \rightarrow \infty$ and $m \in \omega(\log_q(\ell))$ we have
\begin{align*}
\mathcal{R}\sim
\delta^2 \frac{1}{\xi}-\delta\Big(1+\frac{1}{\xi}\Big)+1
\end{align*}
\end{theorem}
\ifShortVersion
\begin{IEEEproof}
See extended version of this paper~\cite{ott2021extended}.
\end{IEEEproof}
\else
\begin{IEEEproof}
From the simplified GV bound given in Theorem~\ref{theo:simpl_GV} it follows that if
\begin{align}
\label{ineq:simpl GV in terms of k}
q^{mk}(d-1)\binom{\ell+d-2}{\ell-1}\gamma_q^\ell q^{(d-1)(m+\eta-\frac{d-1}{\ell})}\leq q^{mn},
\end{align}
then there is a $\ell$-sum-rank metric code $\Code(n,k,d)$.
We transform \ref{ineq:simpl GV in terms of k} into
\begin{align*}
\frac{k}{n}<&\Big(\frac{d}{n}\Big)^2 \frac{\eta}{m}-\frac{d}{n}\Big(1+\frac{\eta}{m}+\frac{2\eta}{nm}\Big)+1+\frac{1}{n}+\frac{\eta}{nm}+\frac{\eta}{n^2m}\\
&-\frac{\log_q\Big((d-1)\cdot\binom{\ell+d-2}{\ell-1}\Big)}{mn} -\frac{\log_q(\gamma_q)}{\eta m}.
\end{align*}
By substituting $\delta\coloneqq \frac{d}{n}$
we get
\begin{align*}
=&\delta^2 \frac{\eta}{m}-\delta\Big(1+\frac{\eta}{m}+\frac{2\eta}{nm}\Big)+1+\frac{1}{n}+\frac{\eta}{nm}+\frac{\eta}{n^2m}\\
&-\frac{\sum_{i=1}^{\delta n-1}\log_q\Big(1+\frac{\ell-1}{i}\Big)+\log_q(\delta n-1)}{mn} -\frac{\log_q(\gamma_q)}{\eta m}\\
&=:\mathcal{R}_*.
\end{align*}
Let $\xi$ be a constant. Consider the following limit:
\begin{align*}
\lim\textstyle_{\stackrel{m=\eta\xi \rightarrow \infty}{m \in \omega(\log_q(\ell))}}\mathcal{R}_*=
\delta^2 \frac{1}{\xi}-\delta\Big(1+\frac{1}{\xi}\Big)+1.
\end{align*}
\end{IEEEproof}
\fi

\subsection{Numerical Comparison}

We compare our simplified and asymptotic bounds to the exact bounds in two parameter regimes/asymptotic settings:
\begin{itemize}
\item \emph{Bounded Block Size}: We keep the extension degree $m$ and the block size $\eta$ constant, and let the number of blocks $\ell$ go to infinity. This is the case for which there are already asymptotic bounds, see~\cite{byrne2020fundamental}.
\item \emph{Growing Block Size}: We let all parameters $\eta,m,\ell$ grow to infinity proportionally. For the plots with finite parameters, we choose $\tfrac{\ell}{\eta}$ and $\tfrac{\eta}{m}$ to be constants close to~$1$. 
\end{itemize}
In contrast to \cite{byrne2020fundamental}, we are able to compare all the bounds for quite large parameters ($n=2^{11}$ and even more) since we use the efficient algorithms for computing the sum-rank ball size from \cite{puchinger2020generic}.

\subsubsection{Bounded Block Size}

\begin{figure}[ht!]
\begin{center}
\input{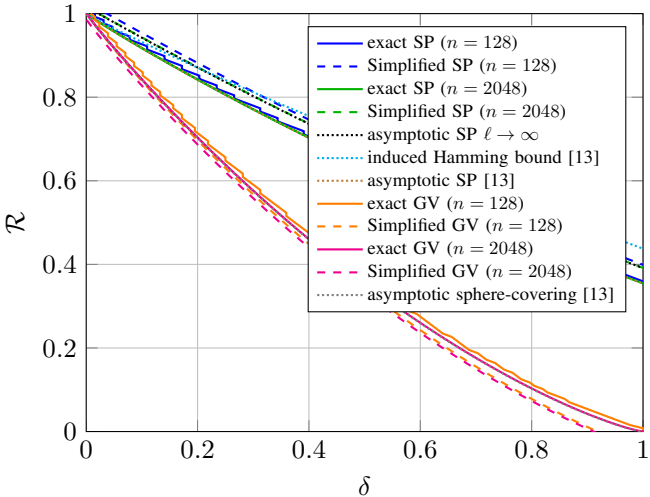}
\end{center}
\vspace{-0.5cm}
\caption{comparison of different bounds for fixed value $q=2$ $\eta=8$ $m=16$ for different values of $n$ ($\ell=\frac{n}{\eta}$)  
}
\label{fig:asy_type1}
\end{figure}

In Figure~\ref{fig:asy_type1}, we compare our simplified bounds (cf.\ Lemma~\ref{Vol_lower_bounds} and \ref{theo:simpl_GV}) with the corresponding exact bounds (cf.\ Theorem~\ref{SP_bound} and \ref{GV-Bound}) and for the SP bound additionally with the asymptotic bound given in Theorem~\ref{Theo:asy_SP} (ii). Moreover we compare our SP and GV bounds to the asymptotic induced Hamming bound and the asymptotic SP and sphere-covering bound, given in \cite[Theorem 4.4 and Corollary 4.10]{byrne2020fundamental}.
The simplified bounds are further away from the exact bounds for this parameter regime (compared to the ``growing block size'' case), since the bounds on $\VolB(t)$ are better suited for $\ell \approx \eta$.
For $n=2^{11}$ the simplified and the asymptotic SP bounds are nearly identical.  From $\delta \geq 0.2$ the asymptotic SP bound is closer to the exact bound than the induced Hamming bound for this setting of parameters.
The asymptotic sphere-packing and sphere-covering bound \cite[Corollary 4.10]{byrne2020fundamental} nearly match the exact SP and GV bound (for $n=2^{11}$), respectively. 

\subsubsection{Growing Block Size}
\begin{figure}[ht!]
\begin{center}
\input{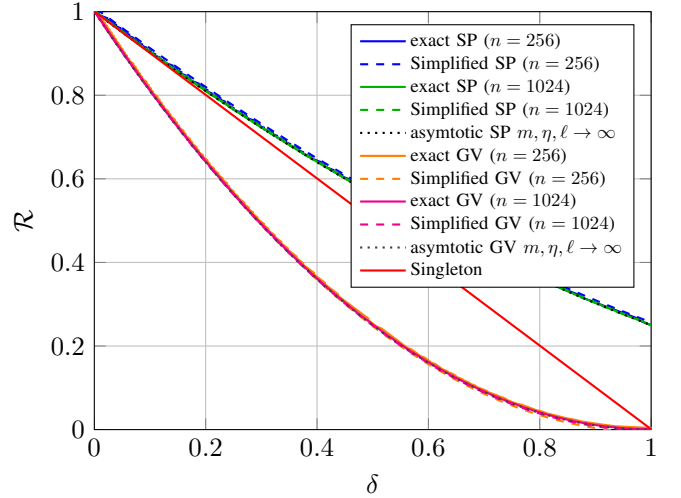}
\end{center}
\vspace{-0.5cm}
\caption{comparison of different bounds for fixed value of $q=2$ and different values of $n$ with $\eta=\ell$ and $m=2\cdot\eta$}
\label{fig:asy_type2_rho=1}
\end{figure}

In Figure~\ref{fig:asy_type2_rho=1}, 
we compare the asymptotic bounds given in Theorem~\ref{Theo:asy_SP} (i) and Theorem~\ref{Theo: asy GV} with the corresponding exact bounds (cf.\ Theorem~\ref{SP_bound} and \ref{GV-Bound}) and with the simplified bounds (cf.\ Lemma~\ref{Vol_lower_bounds} and \ref{theo:simpl_GV}) for two different parameter sets $\ell$, $m$, $\eta$ and $n$.
One can see that for this asymptotic setting the simplified and the exact bounds move closer together for growing $\eta$, $\ell$.
For $n= 2^{10}$ there is no significant difference between the simplified and the exact bounds.
Furthermore, the bounds for this finite $n$ almost coincide with the asymptotic bounds.

\section{Genericity Results}

In this section, we derive the two genericity results. We start with a statement, that codes attaining almost the GV bound are the generic case.

\subsection{Random Linear Codes almost attain the GV bound with high probability}

\begin{theorem}
For $q,m,n,d$, choose $\epsilon \in \Big(0,1-\log_q\big(\VolB(d-1)^{\frac{1}{mn}}\big)-n^{-1}\Big]$ and $k\coloneqq n(1-\log_{q}(\VolB(d-1)^{\frac{1}{mn}})-\epsilon)$.
Let $\Code$ be chosen uniformly at random from the set of linear codes length $n$ and dimension $k$ over $\Fqm$.
Then, $\Code$ has minimum distance $\geq d$ with probability at least $1-e^{-\Omega(mn)}$.
\end{theorem}

\begin{IEEEproof}
Instead of drawing a code uniformly at random from the set of codes with dimension exactly $k$, we consider the following random choice:
Choose a matrix $\G\in \Fqm^{k\times n}$ by drawing its entries independently uniformly at random, and take its row space as the code $\Code$.
Let $A$ be the event that $\rk(\G) =k$ and $B$ be the event that the minimum sum-rank distance of the code is $\geq d$.
The sought probability of the claim is then given by the conditional probability $P(B \mid A)$, since the event $A$ corresponds to all codes of dimension exactly $k$.

First note that
$P(\neg A)
=\sum_{\x\in\Fqm^k\setminus\{\0\}}P(\x\cdot\G=\0)
=\sum_{\x\in\Fqm^k\setminus\{\0\}}\frac{1}{q^{mn}}
=(q^{mk}-1)\cdot \frac{1}{q^{mn}}
<e^{-\Omega(mn)}.$

We also bound the probability $P(B)$.
For a given $\info\in \Fqm^k$, denote by $E_\info$ the event that $\wtsr(\info\cdot \G)<d$.
Then, by the union bound, we have
\begin{equation*}
P(\neg B) = P(\textstyle\bigcup_{\info \in \Fqm^{k}} E_\info)\leq \textstyle\sum_{\info \in \Fqm^{ k}}P(E_\info).
\end{equation*}
For $\info = \0$, we have $P(E_\info)=0$ and for $\info \neq 0$, we get
\begin{align*}
P(E_\info)&=\textstyle\sum_{\c \in \Fqm^{ n} \wtsr(\c)\leq d-1}P(\info\cdot \G = \c)\\
&=\textstyle\sum_{\c \in \Fqm^{n} \wtsr(\c)\leq d-1}\frac{1}{q^{m\cdot n}}
= \frac{\VolB(d-1)}{q^{m\cdot n}}.
\end{align*}
Hence, we can bound
\begin{equation*}
P(B) \geq 1-\tfrac{q^{m\cdot k}\cdot\VolB(d-1)}{q^{m\cdot n}} \geq 1- q^{-mn\epsilon}
\end{equation*}
for the given choice of $k$ and $\epsilon$.
The union bound implies
\begin{equation*}
P(B \mid A) \geq P(B \cap A) \geq 1-e^{-\Omega(mn)},
\end{equation*}
which proves the claim.
\end{IEEEproof}
\subsection{Probability that Random codes are MSRD}

In the following, we derive two lower bounds on the probability that a random linear code is MSRD.
The two bounds are adaptions of the two bounds given by Neri et al.\ in \cite{neri2018genericity} for the rank metric ($\ell=1$), to the general case.
As in \cite{neri2018genericity}, we use the Schwartz--Zippel Lemma together with a counting argument on matrices. The difference to Neri et al.'s proof is that these matrices have a special block structure, which results in bounds that interpolate smoothly between the Hamming and rank case.
It is interesting to note that, in contrast to the two bounds in \cite{neri2018genericity}, the two bounds are advantageous over the other in different parameter ranges. This is due to the nature of the used bounds on the number of these matrices.

Recall that the Schwartz--Zippel Lemma states that, for a non-zero polynomial $f\in \Fqm[x_1, \ldots, x_r]$ of degree $d\geq 0$ and independently uniformly distributed random variables $v_1, \ldots, v_r$ over a subset $\mathcal{F}$ of $\Fqm$ the following probability bound holds: $Pr(f(v_1, \ldots, v_r)=0)\leq\frac{d}{|\mathcal{F}|}$.

We start with a characterization of a code being MSRD (Lemma~\ref{lem:MSRD} below) and use the following notation.

\begin{notation}
We denote by $\mathcal{A}_{\ell,t}$ and $\mathcal{U}_{\ell,t}$ the following sets of block matrices.
\begin{align*}
\mathcal{A}_{\ell,t}\coloneqq &\Big\{\A=\bigoplus\textstyle_{i=1}^\ell \A_i
\in \Fq^{t\times n}\mid  \A_i\in \Fq^{t_i\times \eta},\\ &\rk(\A_i)=t_i, \textstyle\sum_{i=0}^{\ell}t_i=t\Big\}, \quad  \forall t\in \{1, \ldots, n\}.
\\
\mathcal{U}_{\ell,t}\coloneqq&\Big\{\U=\bigoplus_{i=1}^\ell U_i \in \mathcal{A}_{\ell,t}\mid \U_i \text{ upper triangular matrix}\Big\}.
\end{align*}
\end{notation}

In the following lemma, the equivalence (i)$\Leftrightarrow$(ii) was already studied in a similar form in \cite{martinez2019universal,almeida2020systematic,puchinger2020generic}.

\begin{lemma}
\label{lem:MSRD}
Let $\Code(n,k,d)$ be a linear sum-rank metric code with parity check matrix $\H\in  \Fqm^{n-k\times n}$ and generator matrix $\G\in  \Fqm^{k\times n}$.
The following statements are equivalent:
\begin{itemize}
\item[(i)]$\Code$ is MSRD
\item[(ii)]$\rk_{\Fqm}(\A  \G^\top)=k\quad \forall \A\in \mathcal{A}_{\ell, k}$
\item[(iii)]$\rk_{\Fqm}(\U \G^\top)=k\quad \forall \U\in \mathcal{U}_{\ell, k}$.
\end{itemize}
\end{lemma} 
\ifShortVersion
\begin{IEEEproof}
See extended version of this paper~\cite{ott2021extended}.
\end{IEEEproof}
\else
\begin{IEEEproof}
The equivalence (i)$\Leftrightarrow$(ii) follows directly from \cite{martinez2019universal,almeida2020systematic,puchinger2020generic}.
For (ii)$\Leftrightarrow$(iii), since $\mathcal{U}_{\ell, k}\subset \mathcal{A}_{\ell, k}$ it is sufficient to show, that (iii)$\Rightarrow$(ii). Therefore we assume, that $\rk_{\Fqm}(\U \G^\top)=k\quad \forall \U\in \mathcal{U}_{\ell, k}$.
Let  $\A=\bigoplus_{i=1}^\ell \A_i \in\mathcal{A}_{\ell, k}$. Then
$\A_i\in \Fq^{t_i\times \eta}$ 
with $\rk(\A_i)=t_i \quad  \forall i\in \{1, \ldots, \ell\}$ and $\sum_{i=0}^{\ell}t_i=t$.
For each $i\in \{1, \ldots, \ell\}$ let $\U_i$ be the reduced echolon form of $\A_i$, then there is a regular matrix $\X_i \in \Fq^{t_i\times t_i}$ with $\A_i=\X_i \U_i$.
We define $\X\coloneqq \bigoplus_{i=1}^{\ell}\X_i$ and $\U\coloneqq \bigoplus_{i=1}^{\ell}\U_i$, i.e. $\X$ is regular, $\U \in \mathcal{U}_{\ell, k}$ and $\A=\X \U$.
Since $\X_i$ has full rank, it holds
$
\rk_{\Fqm}(\A \G^\top)=\rk_{\Fqm}(\X \U \G^\top)=\rk_{\Fqm}(\U \G^\top)=k.
$
\end{IEEEproof}
\fi
We also derive upper bounds on the cardinality $|\mathcal{A}_{\ell, t}|$ and $|\mathcal{U}_{\ell, t}|$.

\begin{lemma}
\label{cardinality_A_ell,t}
For the cardinality of $\mathcal{A}_{\ell, t}$ it holds 
\begin{align*}
|\mathcal{A}_{\ell, t}|
=\sum_{t\in \tau_{t,\ell,\mu}} \prod\textstyle_{j=0}^{t_i-1}(q^\eta-q^j) 
\leq \binom{t+\ell-1}{\ell-1}q^{\eta t}.
\end{align*}
\end{lemma}
\ifShortVersion
\begin{IEEEproof}
See extended version of this paper~\cite{ott2021extended}.
\end{IEEEproof}
\else
\begin{IEEEproof}
Since the number of matrices $\A_i\in \Fq^{\eta\times t_i }$ of rank $t_i$ is 
\[
\prod\textstyle_{j=0}^{t_i-1}(q^\eta-q^j)\leq q^{\eta t_i} \quad\forall i =1,\ldots,\ell,
\]
it follows, that for a fixed weight decomposition $t=\sum_{i=1}^\ell t_i$ with the restriction, $0\leq t_i \leq \eta$ the number of matrices 
$\A=\A_1\oplus\A_2\oplus\ldots \oplus \A_\ell
\in \Fq^{n\times t}$ with $\A_i\in \Fq^{\eta\times t_i }$ and $\rk(\A_i)=t_i$ is $\prod_{i=1}^\ell \prod_{j=0}^{t_i-1}(q^\eta-q^j)\leq   q^{\eta \sum_{i=1}^\ell t_i}=q^{\eta t}$.
With the number of ordered partitions $\tau_{t,\ell,\mu}$ and its upper bound (see (\ref{ineq: tau})) we get
\begin{equation*}
|\mathcal{A}_{\ell, t}|
=\textstyle\sum_{t\in \tau_{t,\ell,\mu}}\prod_{i=1}^\ell \prod_{j=0}^{t_i-1}(q^\eta-q^j) \leq \binom{t+\ell-1}{\ell-1}q^{\eta t}.
\end{equation*}
\end{IEEEproof}
\fi

\begin{lemma}
\label{cardinality_U_ell,t}
For the cardinality of $\mathcal{U}_{\ell, t}$ it holds 
\begin{align*}
|\mathcal{U}_{\ell, t}|
&=\textstyle\sum_{t\in \tau_{t,\ell,\mu}}\prod_{i=1}^\ell \begin{bmatrix}\eta\\t_i\end{bmatrix}_{q} \leq \binom{t+\ell-1}{\ell-1}q^{t(\eta-\frac{t}{\ell})}\cdot \gamma_q^{\ell}.
\end{align*}
\end{lemma}
\ifShortVersion
\begin{IEEEproof}
See extended version of this paper~\cite{ott2021extended}.
\end{IEEEproof}
\else
\begin{IEEEproof}
The number of upper triangular matrices $\U_i\in \Fq^{t_i\times \eta}$ of rank $t_i$
is equal to the number of $t_i$-dimensional subspaces of $\Fq^\eta$ and therefore equal to the $q$-binomial coefficient 
$\begin{bmatrix}\eta\\t_i
\end{bmatrix}_{q}$. With the same arguments as in the proof of Lemma~\ref{cardinality_A_ell,t} the equality
$
|\mathcal{U}_{\ell, t}|
=\textstyle\sum_{t\in \tau_{t,\ell,\mu}}\prod_{i=1}^\ell \begin{bmatrix}\eta\\t_i\end{bmatrix}_{q}
$
follows.
Using the upper bound 
$
\begin{bmatrix}\eta\\t_i\end{bmatrix}_{q}<\gamma_q\cdot q^{t_i(\eta-t_i)}
$
(see \cite{koetter2008coding}, \cite{berlekamp1980technology}) we get
\begin{align*}
|\mathcal{U}_{\ell, t}|
&<\textstyle\sum_{t\in \tau_{t,\ell,\mu}}\gamma_q^\ell q^{\sum_{i=1}^\ell t_i\eta-t_i^2}\\\
&\leq |\tau_{t,\ell,\mu}|\cdot \gamma_q^\ell q^{t\eta}\cdot\textstyle \max_{\ve{t}\in \tau_{t,\ell,\mu}}q^{-\sum_{i=1}^\ell t_i^2}.
\end{align*}
As derived in \cite[Proof of Theorem 4]{puchinger2020generic} the choice $t_i=\frac{t}{\ell}$ leads to the upper bound
$
\max_{\ve{t}\in \tau_{t,\ell,\mu}}-\Big\{\sum_{i=1}^\ell t_i^2\Big\}\leq-\frac{t^2}{\ell}.
$
Hence
\[
|\mathcal{U}_{\ell, t}|<|\tau_{t,\ell,\mu}|\cdot \gamma_q^\ell q^{t\eta-\frac{t^2}{\ell}}.
\]
With the upper bound on the cardinality of $\tau_{t,\ell,\mu}$ (see (\ref{ineq: tau})) the statement follows.
\end{IEEEproof}
\fi

The lemmas above give the following probability bound.
The bound converges to $1$ for fixed $n,k,\ell$ and $m \to \infty$.
This means that for large enough extension degree $m$, most linear codes are MSRD.

\begin{theorem}
\label{Theo:genericity}
Let $\G$ be a systematic generator matrix $[\I_k \mid \X] \in \Fqm^{k \times n}$, where entries of $\X$ are independently and uniformly chosen from $\Fqm$, and denote by $\Code$ the row space of $\G$.
Then, $\Code$ is an MSRD code (w.r.t.\ $\ell$) with probability at least
\begin{align*}
1-k\tbinom{k+\ell-1}{\ell-1}q^{\eta k-m}.
\end{align*}
\end{theorem}
\begin{IEEEproof}
From Lemma~\ref{lem:MSRD} we know that $\Code$ is MSRD if and only if  
$\rk_{\Fqm}(\G \cdot \A)=k$ for all $\A\in \mathcal{A}_{\ell, k}$. This leads to the fact that  $\Code$ is not MSRD if and only if there exists a matrix $\A\in\mathcal{A}_{\ell, k}$, such that $\det(\G \cdot \A)=0$.
Considering the entries of $\X$ as variables $x_1, \ldots x_{k(n-k)}$,then it holds for the product 
\[
\G\cdot \A= [\I_k|\X]\cdot\A= \Big(a_{ij}+\textstyle\sum_{r=k+1}^n x_{ir-k}a_{rj}\Big)_{ij}
\]
that each variable $x_1, \ldots x_{k(n-k)}$ is contained in at most one row.
Hence, the determinant of this product is a multivariate polynomial $f_\A\coloneqq\det([\I_k|\X]\cdot\A)\in \Fq[x_1, \ldots x_{k(n-k)}]$ of degree at most $k$ for each $\A\in \mathcal{A}_{\ell, k}$.
Using the notation
\[
f(x_1, \ldots x_{k(n-k)})\coloneqq\textstyle\prod_{\A \in \mathcal{A}_{\ell, k}}f_\A(x_1, \ldots x_{k(n-k)})
\]
we have
$P(\Code \text{ is not MSRD})=P(f(x_1, \ldots x_{k(n-k)})=0)$.
Since $f\neq0$ and the variables $x_1, \ldots x_{k(n-k)}$ are independently and uniformly distributed, it follows with the Schwartz--Zippel Lemma that the probability $P(f(x_1, \ldots x_{k(n-k)})=0)\leq\frac{\deg f}{|\Fqm|}=\frac{k|\mathcal{A}_{\ell, k}|}{q^m}$. 
From Lemma~\ref{cardinality_A_ell,t} it follows
\begin{equation*}
|\mathcal{A}_{\ell, k}|
\leq \tbinom{k+\ell-1}{\ell-1}q^{\eta k},
\end{equation*}
which proves the claim.
\end{IEEEproof}
The number of matrices in $\mathcal{U}_{\ell, t}$ is always smaller than the number of matrices in $\mathcal{A}_{\ell, t}$. This motivates the following bound, which uses the criterion of Lemma~\ref{lem:MSRD} (iii) and Lemma~\ref{cardinality_U_ell,t}. 

\begin{theorem}
\label{Theo:genericity_improved}
Let $\G=[\I_k \mid \X]\in \Fqm^{k \times n}$ be a systematic generator matrix, with a random matrix $\X$, i.e., the entries are independently and uniformly chosen from $\Fqm$. The row space of $\G$ is a is a linear $\ell$-sum--rank metric code over $\Fqm$, which we denote by $\Code$.
The probability $p$ that $\Code$ is MSRD is lower bounded by
\[
p\geq 1-k\tbinom{k+\ell-1}{\ell-1}q^{k(\eta-\frac{k}{\ell})-\frac{\ell}{4}-m}\cdot \gamma_q^{\ell}.
\]
\end{theorem}
\ifShortVersion
\begin{IEEEproof}
See extended version of this paper~\cite{ott2021extended}.
\end{IEEEproof}
\else
\begin{IEEEproof}
With Lemma~\ref{lem:MSRD} (iii) and Lemma~\ref{cardinality_U_ell,t} using the same arguments as in the proof of Theorem~\ref{Theo:genericity} by just replacing $\mathcal{A}_{\ell, k}$ by $\mathcal{U}_{\ell, k}$ one gets
\begin{align*}
P(\Code \text{ is not MSRD})&\leq \frac{k|\mathcal{U}_{\ell, k}|}{q^m}\\
&\leq k\tbinom{k+\ell-1}{\ell-1}q^{k(\eta-\frac{k}{\ell})-\frac{\ell}{4}}\cdot \gamma_q^{\ell}q^{-m}.
\end{align*}
\end{IEEEproof}
\fi
\subsection{Numerical Comparison}

\begin{figure}[h!]
\begin{center}
\input{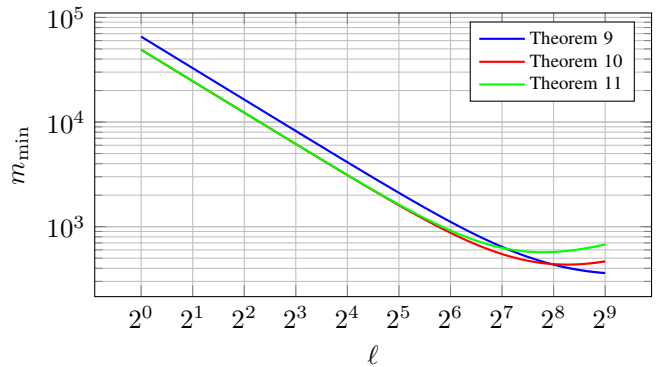}
\end{center}
\vspace{-0.5cm}
\caption{comparison of Theorem~\ref{Theo:genericity} and Theorem~\ref{Theo:genericity_improved} for $n=2^{10}$, $k=2^8$ and $q=4$.}
\label{fig:genericity}
\end{figure}

For the bounds given in Theorem~\ref{Theo:genericity} and \ref{Theo:genericity_improved}, Figure~\ref{fig:genericity} shows the minimal values of the extension degree $m$, for which there is a non-zero probability that a code is MSRD for varying numbers of blocks $\ell$ and constant values of $n$, $k$ and $q$.
These minimal values $m_{\min}$ smoothly interpolate between the known extreme cases: $\ell=1$ (cf. \cite{neri2018genericity}) and $\ell=n$ (cf. \cite{ravagnani2019sparsity}). 
Since the complementary probability of the two bounds differ in the factor $\zeta \coloneqq q^{-\frac{k^2}{\ell}-\frac{\ell}{4}}\gamma_q^\ell$ it depends on the relation of $\zeta$ to $1$, which of the two bounds is the better one.
This can be observed in Figure~\ref{fig:genericity}. For small values of $\ell$, the bound derived in Theorem~\ref{Theo:genericity_improved} is better and in turn for large values of $\ell$, Theorem~\ref{Theo:genericity} provides the better bound. 
Hence, both bounds have advantages in certain parameter ranges of $\ell$.
This is different from the bounds in \cite{neri2018genericity}, where the bound in \cite[Theorem 26]{neri2018genericity} is an improvement of the bound derived in \cite[Theorem 21]{neri2018genericity}.

\bibliographystyle{IEEEtran}
\bibliography{main}
\end{document}